\documentclass[letterpaper,12pt,]{article}

\usepackage{color}
\usepackage{hyperref} 
\hypersetup{colorlinks=true,citecolor=blue}

\def\mnras{\emph {MNRAS}}

\def \bea {\begin{eqnarray}}
\def \ena {\end{eqnarray}}                

\def	\cm		{\,{\rm cm}}
\def	\m		{\,{\rm m}}

\def	\B		{\,{\rm B}}

\def    \exp 		{\,{\rm exp}}
\def	\g		{\,{\rm g}}

\def	\K		{{\,\rm K}}

\def	\s		{\,\rm s}
\def	\sp		{\,{\rm sp}}
\def	\ed		{\,{\rm ed}}
\def	\gas	{\,{\rm gas}}
\def	\eff	  {\,{\rm eff}}
\def	\tot	  {{\rm tot}}

\def	\H		{{\rm H}}

\def	\T		 {{\rm T}}

\def	\ba		{{\bf a}}

\def	\bJ		{{\bf J}}

\def	\xhat		{\hat{\bf x}}
\def	\yhat		{\hat{\bf y}}

\def	\FT		{{\rm FT}}

\def	\VRE		{{\rm VRE}}
\def	\vib		{{\rm vib}}
\def	\peak		{{\rm peak}}
\def	\icoll	        {{\rm icoll}}

\font\mib=cmmib10

\def \bmu {{\hbox {\mib\char"16}}}

\newcommand   \Angstrom	  {\,{\rm \AA}}		

\title{Spinning Dust Emission from Wobbling Grains: Important Physical Effects and Implications}
\author{
        Thiem Hoang\\
		Canadian Institute for Theoretical Astrophysics\\
		University of Toronto\\
		60 St. George Street, Toronto, ON M5S 3H8, \underline{Canada}
        \and
        A. Lazarian\\
        Department of Astronomy\\
        University of Wisconsin-Madison\\
        Madison, WI 53705, \underline{USA}
}

\date{\today}

\usepackage{epsfig}
\begin{document}

\maketitle

\begin{abstract}
We review major progress on the modeling of electric dipole emission
from rapidly spinning tiny dust grains, including polycyclic aromatic hydrocarbons (PAHs).
We begin by summarizing the original model of spinning dust
proposed by Draine and Lazarian and recent theoretical results
improving the Draine and Lazarian model. The review
is focused on important physical effects that were disregarded in earlier
studies for the sake of simplicity and recently accounted for
by us, including grain wobbling due to internal relaxation, impulsive excitation
by single-ion collisions, the triaxiality of grain shape,
charge fluctuations, and the turbulent nature of astrophysical environments.
Implications of the spinning dust for constraining physical
properties of ultrasmall dust grains and environment conditions
are discussed. We discuss the alignment of tiny dust grains
and possibility of
polarized spinning dust emission. Suggestions for constraining the
alignment of tiny grains and polarization of spinning dust are
also discussed.

\end{abstract}


\section{Introduction}

Diffuse Galactic microwave emission carries important information on
the fundamental
properties of the interstellar medium, but it also interferes with
Cosmic Microwave Background (CMB) experiments
(see  Bouchet et al.~\cite{Bouchet:1999p4616};
Tegmark et al.~\cite{Tegmark:2000p5597}). Precision cosmology with {\it Wilkinson Microwave 
Anisotropy Probe} 
({\it WMAP}) and Planck satellite requires a good model of the microwave 
foreground emission to allow for reliable subtraction of Galactic contamination 
from the CMB radiation.

The discovery of an anomalous microwave emission (hereafter AME) in
the range from $10-100$ GHz
illustrates well the treacherous nature of dust.
Until very recently it has been thought that
there are three major components of the diffuse Galactic foreground:
synchrotron emission, free-free radiation from plasma (thermal bremsstrahlung)
and thermal emission from dust. In the microwave range, the latter
is subdominant, leaving essentially two
components. However, it is exactly in this range that an anomalous
emission component was reported (Kogut et al.~\cite{Kogut:1996p5303} and \cite{Kogut:1996p5293}). In the
paper by de Oliveira-Costa et al.~\cite{deOliveiraCosta:2002p4672} this emission was nicknamed
``Foreground X'', which properly reflects its mysterious nature.
This component is spatially correlated with 100 $\mu$m thermal
dust emission, but its intensity is much higher than one would expect
by directly extrapolating the thermal dust emission spectrum
to the microwave range.

An early explanation for AME was proposed by Draine \& Lazatian~\cite{Draine:1998p763}
and \cite{Draine:1998p784} (hereafter DL98 model), where it was identified as
electric dipole emission 
from very small grains (mostly containing polycyclic aromatic hydrocarbons--PAHs) that spin
rapidly due to several processes, including gas-grain interactions and dust
infrared emission.
Although spinning dust emission had
been discussed previously (see Erickson~\cite{Erickson:1957p4806};
Ferrera \& Dettmar~\cite{Ferrara:1994p4811}), Draine \& Lazarian
were the first to include the variety of excitation and damping processes that are 
relevant for very small grains.

While the DL98 model appears to be in general agreement with observations 
(see \cite{Lazarian:2003p715}; \cite{Finkbeiner:2004p471}), it did not 
account for some important effects, namely, the non-sphericity of grain 
shapes, internal relaxation within grain, and transient spin-up due to ion collisions. 

This induced more recent work in order to improve the original DL98 model.
The recent papers include (Ali-Ha{\"\i}moud et al.~\cite{AliHaimoud:2009p641}; Hoang et al.~\cite{Hoang:2010p2383}; 
Ysard~\cite{Ysard:2010p584}; Hoang et al.~\cite{Hoang:2011p2326}; Silsbee et al.~\cite{Silsbee:2011p5567}). 
In this paper we review both the original DL98 model and the ways that it
has been improved recently. We focus on the improvement of dynamics of PAHs
and important physical effects associated with these ultrasmall grains.
Recent reviews of the subject include
Draine \& Lazarian~\cite{1999ASPC..181..133D} and \cite{2002AIPC..609...32L}, and
Lazarian \& Finkbeiner~ \cite{2003NewAR..47.1107L}.

In \S 2 we briefly present the history of AME and discuss the original DL98
model including their basic assumptions.
\S 3 presents our principal results improving the DL98 model
from Hoang et al.~\cite{Hoang:2010p2383} and \cite{Hoang:2011p2326}.
From \S 4 to \S 6 we review on grain rotational dynamics and discuss our general approach
to calculate power spectrum of spinning dust emission,
grain angular momentum distribution and emissivity for PAHs of arbitrary shapes.
In \S 7, we discuss the implications of spinning dust for constraining physical parameters
of PAHs as well as environment conditions.
The possibility of polarization of spinning dust and its constraint is discussed in \S 8.
A summary of the present review is given in \S 9.

\section{The original DL98 model}

\subsection{Anomalous microwave emission and PAHs}

The emission spectrum of diffuse interstellar dust was mostly obtained
by the {\it InfraRed Astronomy Satellite} (IRAS) and infrared
spectrometers on the {\it COsmic Background Explorer} (COBE) and on
the {\it InfraRed Telescope in Space} (IRTS).
The emission at short wavelength ($\lambda <50\mu$m)
arises from transiently heated ultrasmall grains (e.g., PAHs). These grains have
such a small heat capacity that the absorption of a single ultraviolet (UV) starlight
photon ($\sim 6$ eV) raises their temperature to $T_{\vib}>200$ K. Typically, these grains
have less than $300$ atoms and can be viewed as large molecules
rather than dust particles. They are, however, sufficiently numerous
to account for most of the prominent $2175$\AA~ absorption feature, and for $\sim 35\%$ of
the total starlight absorption (see e.g. Li \& Draine~\cite{2001ApJ...554..778L}).

The thermal (vibrational) emissivity of these grains is thought to be
negligible at low frequency, because they spend most of their time
cold and only emit most of their energy when they are hot.
These ultrasmall grains (PAHs) are invoked in the DL98 model to account
for the anomalous microwave emission (AME) that was measured in observations.

The first detection of anomalous dust-correlated emission by COBE
(Kogut et al.~\cite{Kogut:1996p5303}, \cite{Kogut:1996p5293}) was
quickly followed by detections in the data sets from Saskatoon (de Oliveira-Costa et al.~\cite{1997ApJ...482L..17D}), OVRO (Leitch et al.~\cite{Leitch:1997p7359}), the 19~GHz survey (de Oliveira-Costa et al.~\cite{deOliveiraCosta:1998p4707}), and
Tenerife (\cite{deOliveiraCosta:1999p4706}).
Initially, AME was identified as thermal bremsstrahlung
from ionized gas correlated with dust (Kogut et al.~\cite{Kogut:1996p5303}) and presumably
produced
by photoionized cloud rims (McCullough et al.~\cite{1999ASPC..181..253M}). This idea was
scrutinized in Draine \& Lazarian~\cite{Draine:1998p763} and criticized on
energetic grounds. Poor correlation of H$\alpha$ with 100~$\mu$m
emission also argued against the free-free explanation (McCullough et al.~\cite{1999ASPC..181..253M}).
These arguments are summarized in \cite{1999ASPC..181..133D}. Later
\cite{deOliveiraCosta:2000p4667} used Wisconsin H-Alpha Mapper (WHAM)
survey data and established that the free-free emission ``is about an order
of magnitude below Foreground X over the entire range of frequencies
and latitudes where it is detected''. The authors concluded that the
Foreground X cannot be explained as the free-free emission. Additional
evidence supporting this conclusion has come from
a study at 5, 8 and 10~GHz by Finkbeiner et al.~\cite{Finkbeiner:2002p492} of several dark
clouds and HII regions, two of which show a significantly rising
spectrum from 5 to 10 GHz.

The recent Wilkinson Microwave Anisotropy Probe (WMAP) data were used
to claim a lower limit of 5\% for the spinning dust fraction at 23 GHz
(Bennet et al.~\cite{Bennett:2003p4582}).  However, other models of spinning dust are not
ruled out by the WMAP data, and in fact fit reasonably well.
Finkbeiner~\cite{Finkbeiner:2004p471} performed a fit to WMAP data
using a CMB template, a
free-free template (based on H$\alpha$-correlated emission plus hot
gas emission near the Galactic center), a soft synchrotron template
traced by the 408 MHz map, a thermal dust extrapolation (Finkbeiner et al.~\cite{Finkbeiner:1999p6816}) and a spinning dust template consisting of dust column
density times $T^3_{d}$.  This fit results in excellent
$\chi^2/dof$ values (1.6, 1.09, 1.08, 1.05, 1.08) at (23, 33, 41, 61, 94) GHz
and a reasonable spectral shape for the average spinning dust
spectrum. 

This WMAP analysis alone does not rule out the Bennet et al.~\cite{Bennett:2003p4582}
hypothesis of hard synchrotron emission, but when combined with the
Green Bank Galactic Plane survey data (Langston et al.~\cite{2000AJ....119.2801L}) at 8 and
14 GHz, spinning dust appears to provide a much better fit than hard
synchrotron (Finkbeiner et al.~\cite{Finkbeiner:2004p461}). 

Spinning dust emission has recently been reported in a wide range
of astrophysical environments, including general ISM (Gold et al.~\cite{Gold:2009p5268}; \cite{Gold:2011p5261};
Planck Collaboration~\cite{PlanckCollaboration:2011p515}), star forming regions
in the nearby galaxy NGC 6946 (Scaife et al.~\cite{Scaife:2010p569}; \cite{2012ApJ...754...94T}),
and Persus and Ophiuchus clouds (Casassus et al.~\cite{Casassus:2008p3387}; 
Tibbs et al.~\cite{Tibbs:2011p541}). Early Planck results have been
interpreted as showing a microwave emission excess from spinning dust in the Magellanic
Clouds (Bot et al.~\cite{Bot:2010p5582}; Planck Collaboration~\cite{PlanckCollaboration:2011p515}).

\subsection{Basic assumptions}


(i) The smallest PAH particles of a few Angstroms are expected to be planar.
The grain size $a$ is defined as the radius of an equivalent sphere of the same mass.
PAHs are assumed to be planar, disk-like with height $L$ and radius $R$ for $a<a_{2}$ and
spherical for $a \ge a_{2}$. The value $a_{2}=6 \Angstrom$ is adopted.

(ii) PAHs usually have electric dipole moment $\bmu$ arising
from asymmetric polar molecules
or substructures ({\it intrinsic dipole moment}) and from the asymmetric
distribution of grain charge. The latter is shown to be less important.

(iii) The grain spins around its symmetry axis $\ba_{1}$ with angular
momentum $\bJ$ parallel to $\ba_{1}$ and $\bJ$ is isotropically oriented in space.

(iv) For a fixed angular momentum, the spinning grain emits electric
dipole radiation at a {\it unique} frequency mode $\nu$, which is equal to
the rotational frequency, i.e., $\nu=\omega/2\pi$.

(v) A grain in the gas experiences collisions with neutral atoms and ions,
interacts with passing ions (plasma-grain interactions), emits
infrared photons following UV absorption, and emits electric dipole radiation.
All these processes result in the damping and excitation of
grain rotation, i.e., they change grain angular momentum $J$ and velocity $\omega$.

(vi) Due to the excitation of various aforementioned processes, the grain
angular velocity randomly fluctuates and its distribution can be approximated as
the Maxwellian distribution function $f_{\rm Mw}(\omega)$.

(vi) The total emissivity per H atom of
the electric dipole radiation from spinning dust at the frequency $\nu$ is
given by
\bea
\frac{j_{\nu}}{n_\H} =\frac{1}{4\pi}\frac{1}{n_\H}
\int_{a_{\min}}^{a_{\max}} da {dn\over da} 
4\pi \omega^2 f_{\rm Mw}(\omega) 2\pi 
\left(\frac{2\mu_{\perp}^{2}\omega^{4}}{3c^{3}}\right)~~~,
\label{eq:504}
\ena
where $n_{\H}$ is the density of H nuclei, $\mu_{\perp}$
is the electric dipole moment perpendicular to the rotation axis, and 
$dn/da$ is the grain size distribution function with $a$ in
the range from $a_{\min}$ to $a_{\max}$.

\section{Improved Model of Spinning Dust Emission}

Ali-Ha{\"\i}moud et al.~\cite{AliHaimoud:2009p641} revisited
the spinning dust model and presented an 
analytic solution of the Fokker-Planck (FP) equation that describes the rotational
excitation of a spherical grain if the discrete nature of impulses from single-ion
collisions can be neglected.

Hoang et al.~\cite{Hoang:2010p2383} (hereafter HDL10) improved the 
DL98 model by accounting for a number of physical effects.
The main modifications in their improved model of spinning dust
emission are as follows.

(i) Disk-like grains rotate with their grain symmetry axis $\ba_{1}$ not perfectly
aligned with angular momentum $\bJ$. The disaligned rotation of
$\bJ$ with $\ba_{1}$ causes the wobbling of the grain principal axes
with respect to $\bJ$ due to internal thermal fluctuations.

(ii) The power spectrum of a freely spinning grain is obtained
using Fourier transform.

(iii) Distribution function for grain angular momentum $J$ and velocity $\omega$ 
are obtained exactly using the Langevin equation (LE) for
the evolution of $\bJ$ in an inertial coordinate system.

(iv) The limiting cases of fast internal relaxation and no internal relaxation
are both considered for calculations of the angular momentum distribution
and emissivity of spinning dust.

(v) Infrequent collisions of single ions which deposit an angular 
momentum larger than the grain angular momentum prior to the collision
are treated as Poisson-distributed events.

The wobbling disk-like grain has anisotropic rotational damping and excitation.
Such an anisotropy can increase the peak emissivity by a factor $\sim 2$, and
increases the peak frequency by a factor $1.4-1.8$, 
compared to the results from the DL98 model. 

The effects of grain wobbling on electric dipole emission was independently
studied in Silsbee et al.~\cite{Silsbee:2011p5567} using the FP equation approach,
but they disregarded the transient spin-up
by infrequent single-ion collisions, and considered two limiting cases
of dust grain temperature $T_{d}\rightarrow 0$ and
$T_{d} \rightarrow \infty$. 

Further improvements of the DL98 model were performed in Hoang et al.~\cite{Hoang:2011p2326},
where a couple of additional effects were taken into account:

(i) emission from very small grains of triaxial ellipsoid ({\it irregular}) shape with
the principal moments of inertia $I_{1}\ge I_{2}\ge I_{3}$.

(ii) effects of the orientation of dipole moment $\bmu$ within grain body
for different regimes of internal thermal fluctuations.

(iii) effects of compressible turbulence on the spinning dust emission.

The work found that a freely rotating irregular grain with a given angular momentum 
radiates at multiple frequency modes. The resulting spinning dust spectrum
has peak frequency and emissivity increasing with the degree of grain shape
irregularity, which is defined by $I_{1}:I_{2}:I_{3}$. Considering the transient heating of
grains by UV photons, the study found that the spinning dust 
emissivity for the case of strong thermal fluctuations is less sensitive to the orientation of 
$\bmu$ than in the case of weak thermal fluctuations. In addition, the emission in a turbulent medium 
increases by a factor from $1.2$--$1.4$ relative to that in a uniform medium, as sonic Mach number $M_{\s}$ increases from $2$--$7$. The latter Mach numbers are relevant to cold phases of
the ISM (see Hoang et al.~\cite{Hoang:2011p2326} for more details).

\section{Grain Rotational Configuration and Power Spectrum}

A discussion of the basic physical processes involved in spinning dust
can be found 
in the review by Yacine Ali-Ha\"{i}moud, which can be found in the same volume.
There, the use of Fokker-Planck equation for describing grain dynamics is
discussed. Here we discuss our numerical approach based
Fourier transform and the Langevin equation, which exhibits a number
of advantages to the FP equation when numerical studies of
grain dynamics are performed and arbitrary shape of PAHs is considered.
We summarize a general
approach to find the spinning dust emissivity from grains of triaxial
ellipsoid shape with $I_{1}>I_{2}>I_{3}$ subject to fast internal relaxation.

\subsection{Torque-free motion and internal relaxation}
\begin{figure}
\begin{center}
\includegraphics[width=0.6\textwidth]{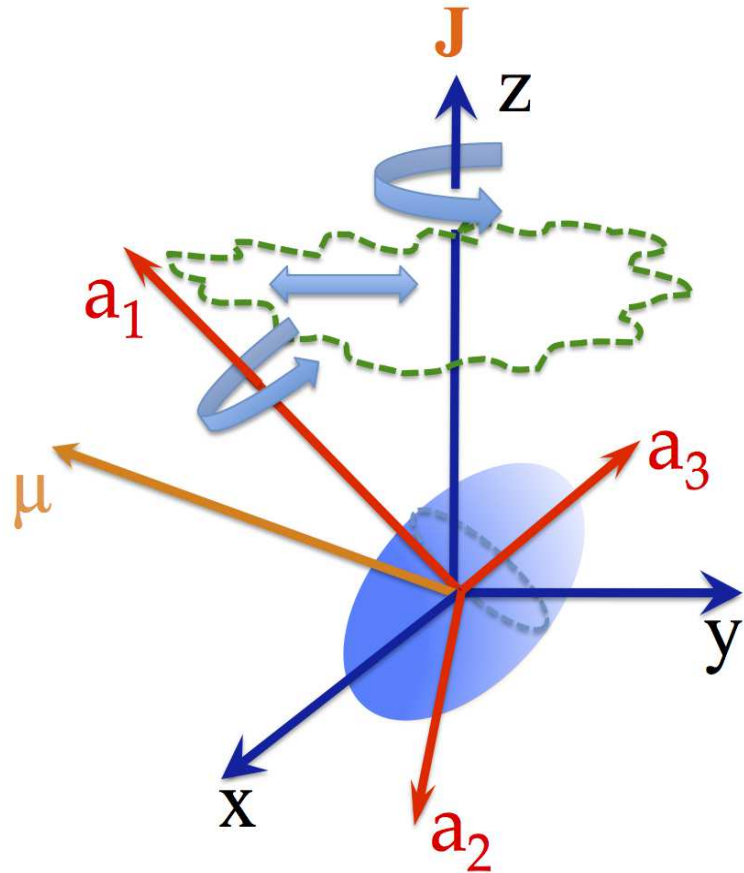}
\caption{Rotational configuration of a triaxial ellipsoid characterized
by three principal axes $\ba_{1},~\ba_{2}$
and $\ba_{3}$ in the inertial coordinate system $xyz$. Grain angular momentum
$\bJ$ is conserved in the absence of external torques and directed
along $z$-axis. The torque-free motion
of the triaxial grain comprises the rotation around the axis of major inertia $\ba_{1}$,
the precession of $\ba_{1}$ around $\bJ$, and the wobbling of $\ba_{1}$
with respect to $\bJ$. The dipole moment $\bmu$, which is fixed to grain body,
moves together with the grain and thus radiates electric dipole emission.}
\label{fig:3Dgrain}
\end{center}
\end{figure}

The dynamics of a triaxial ({\it irregular}) grain is more
complicated than that of a disk-like
grain with $I_{2}=I_{3}$. Indeed, in addition to the precession of
the axis of major inertia $\ba_{1}$
around $\bJ$ as in the disk-like grain, the axis $\ba_{1}$  wobbles rapidly,
resulting in the variation of the angle $\theta$ between $\ba_{1}$ and $\bJ$
(see Figure \ref{fig:3Dgrain}).

To describe the torque-free motion of an irregular grain having a rotational energy
$E_{\rm rot}$, the conserved quantities are taken, including the angular momentum $\bJ$,
and a dimensionless parameter that characterizes the deviation of the grain 
rotational energy from its minimum value,
\bea
q=\frac{2I_{1}E_{\rm rot}}{J^{2}}.\label{eq:603}
\ena

The orientation of the triaxial grain in the lab system is completely described
by three Euler angles $\psi,~\phi$ and $\theta$ (see e.g., Hoang et al.~\cite{Hoang:2011p2326}).
Following \cite{Weingartner:2003p3452}, we define the total number of states $s$ 
in phase space for $q$ ranging from $1$ to $q$ as 
\bea
s\equiv 1-\frac{2}{\pi}\int_{0}^{\psi_{1}}d\psi \left[\frac{I_{3}(I_{1}-I_{2}q)+
I_{1}(I_{2}-I_{3})\cos^{2}\psi}{I_{3}(I_{1}-I_{2})+I_{1}(I_{2}-I_{3})
\cos^{2}\psi}\right]^{1/2},~~~~\label{eq:604}
\ena
where
\bea
\psi_{1}=\cos^{-1}\left[\frac{I_{3}(I_{2}q-I_{1})}{I_{1}(I_{2}-I_{3})}\right]^{1/2},
\ena
for $q>q_{\sp}$ and $\psi_{1}=\pi/2$ for $q\leq q_{\sp}$, with 
$q_{\sp}\equiv I_{1}/I_{2}$ being the separatrix between the two regimes.

The Intramolecular 
Vibrational-Rotational Energy Transfer process (IVRET) due to imperfect elasticity
occurs on a timescale $10^{-2}$ s,
for a grain of a few Angstroms (Purcell~\cite{Purcell:1979p2611}),
which is shorter than the IR emission time. So, when the vibrational energy 
decreases due to IR emission, as long as the Vibrational-Rotational (V-R)
energy exchange exists, interactions between vibrational
and rotational systems maintain a thermal equilibrium, i.e.,
$T_{\rm rot}\approx T_{\vib}$. As a result, the LTE distribution
function of rotational energy reads (hereafter VRE regime;
see Lazarian \& Roberge~\cite{Lazarian:1997p2598}):
\bea
f_{\VRE}(s,J)\propto {\exp}\left(-\frac{E_{\rm rot}}{k_{\B}T_{\rm rot}}\right)
\approx{\exp}\left(-\frac{E_{\rm rot}}{k_{\B}T_{\vib}}\right).\label{eq:605}
\ena

Substituting $E_{\rm rot}$ as a function of $J$ and $q$ from Equation (\ref{eq:603})
into Equation (\ref{eq:605}),  
the distribution function for the rotational energy becomes
\bea
f_{\VRE}(s,J)= A{\exp}\left(-\frac{q(s)J^{2}}{2I_{1}k_{\B}T_{\vib}}\right),
\label{eq:606}
\ena
where $A$ is a normalization constant such that $\int_{0}^{1} f_{\VRE}(s,J)ds=1$.

\subsection{Power spectrum of a freely spinning grain}

Consider a grain with a dipole moment $\bmu$ fixed in the grain body
rotating with an angular momentum $\bJ$. If the grain only spins around
its symmetry axis, then the rotating dipole moment emits radiation at a unique frequency
$\nu$ equal to the rotational frequency, i.e., $\nu=\omega/2\pi$ (see DL98).
The power spectrum for this case is simply a Delta function $\delta(\nu-\omega/2\pi)$
with a unique frequency mode.

For an irregular grain of triaxial ellipsoid shape, the grain rotational dynamics is
more complicated. In general, one can also obtain analytical expressions for
power spectrum, but it is rather tedious.
To find the power spectrum of a freely rotating irregular grain,
Hoang et al.~\cite{Hoang:2010p2383} and \cite{Hoang:2011p2326})
have employed a more simple, brute force approach based on the Fourier transform approach.
First, they represent the dipole moment $\bmu$
in an inertial coordinate system, and then compute its second derivative. 
We obtain
\bea
\ddot{\bmu}=\sum_{i=1}^{3}\mu_{i}\ddot{\ba}_{i},\label{eq:607}
\ena
where $\mu_{i}$ are components of $\bmu$ along principal axes $\ba_{i}$,
$\ddot{\ba}_{i}$ are second 
derivative of $\ba_{i}$ with respect to time, and $i=1, 2$ and $3$. 

The instantaneous emission power by the rotating dipole moment is equal to 
\bea
P_{\ed}(J,q,t)=\frac{2}{3c^{3}}\ddot\bmu^{2}.\label{eq:608}
\ena

The power spectrum is then obtained from the Fourier 
transform (FT) for the components of $\ddot\bmu$. 
For example, the amplitude of $\ddot\mu_{x}$ at the frequency $\nu_{k}$ is 
defined as
\bea
\ddot\mu_{x,k}=\int_{-\infty}^{+\infty} \ddot\mu_{x}(t) 
{\exp}\left(-i2\pi\nu_{k}t\right)dt,
\ena
where $k$ denotes the frequency mode. The emission power at the positive 
frequency $\nu_{k}$ is given by
\bea
P_{\ed,k}(J,q)=\frac{4}{3c^{3}}(\ddot\mu_{x,k}^{2}+\ddot\mu_{z,k}^2+
\ddot\mu_{z,k}^2),
\ena
where the factor $2$ arises from the positive/negative frequency symmetry 
of the Fourier spectrum. To reduce the spectral leakage in the FT, we convolve
the time-dependent function $\ddot\bmu$ with the Blackman-Harris window function
(Harris 1978). The power
spectrum then needs to be corrected for the power loss due to the window function.

The total emission power from all frequency modes for a given $J$ and $q$ then becomes
\bea
P_{\ed}(J,q)=\sum_{k}P_{\ed,k}(J,q)
\equiv\frac{1}{T}\int_{0}^{T}dt\left(\frac{2}{3c^{3}}\ddot\bmu^{2}\right),~~~\label{eq:609}
\ena
where $T$ is the integration time. \footnote{This is the result of Parseval's
Theorem.}

\begin{figure*}
\includegraphics[width=0.5\textwidth]{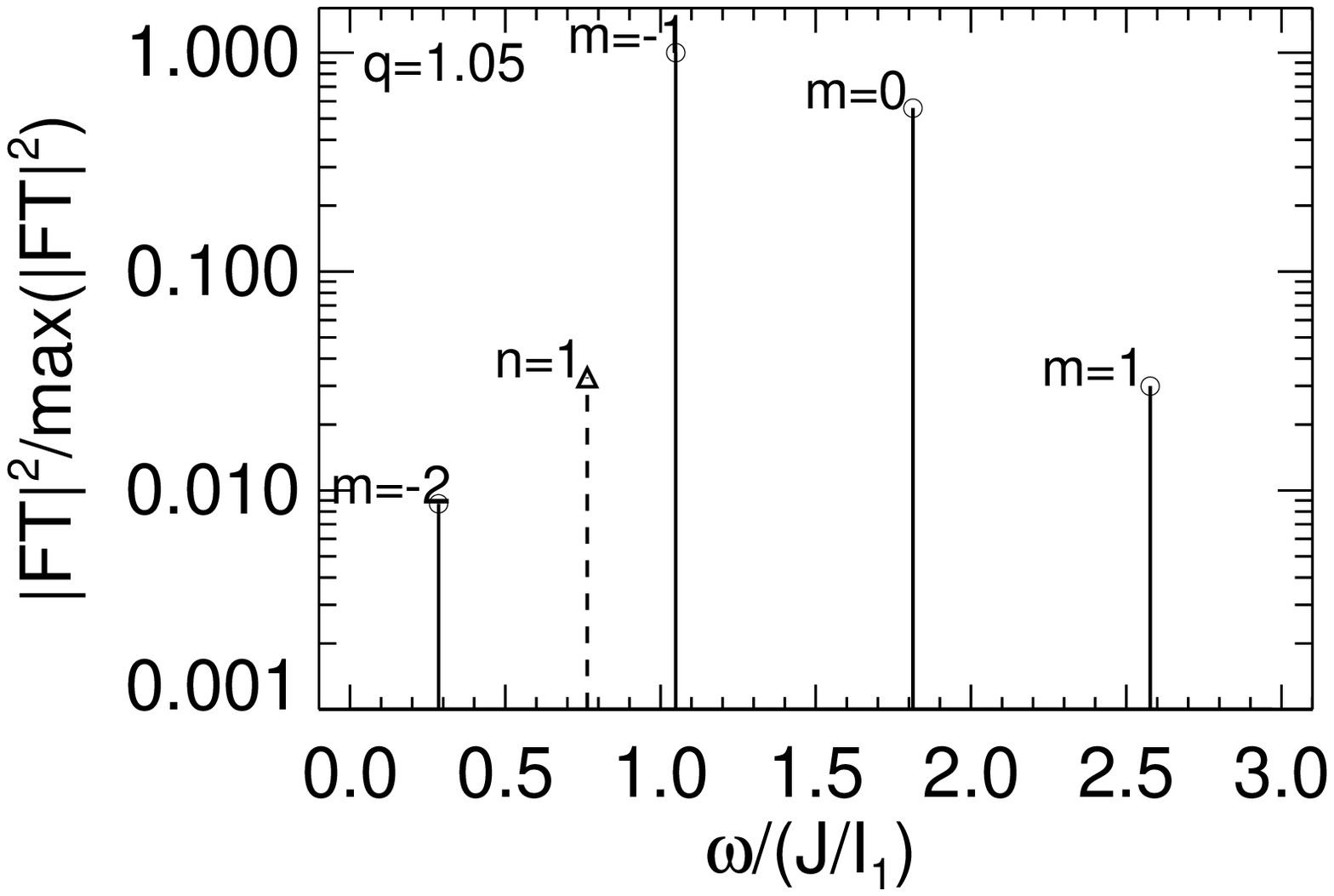}
\includegraphics[width=0.5\textwidth]{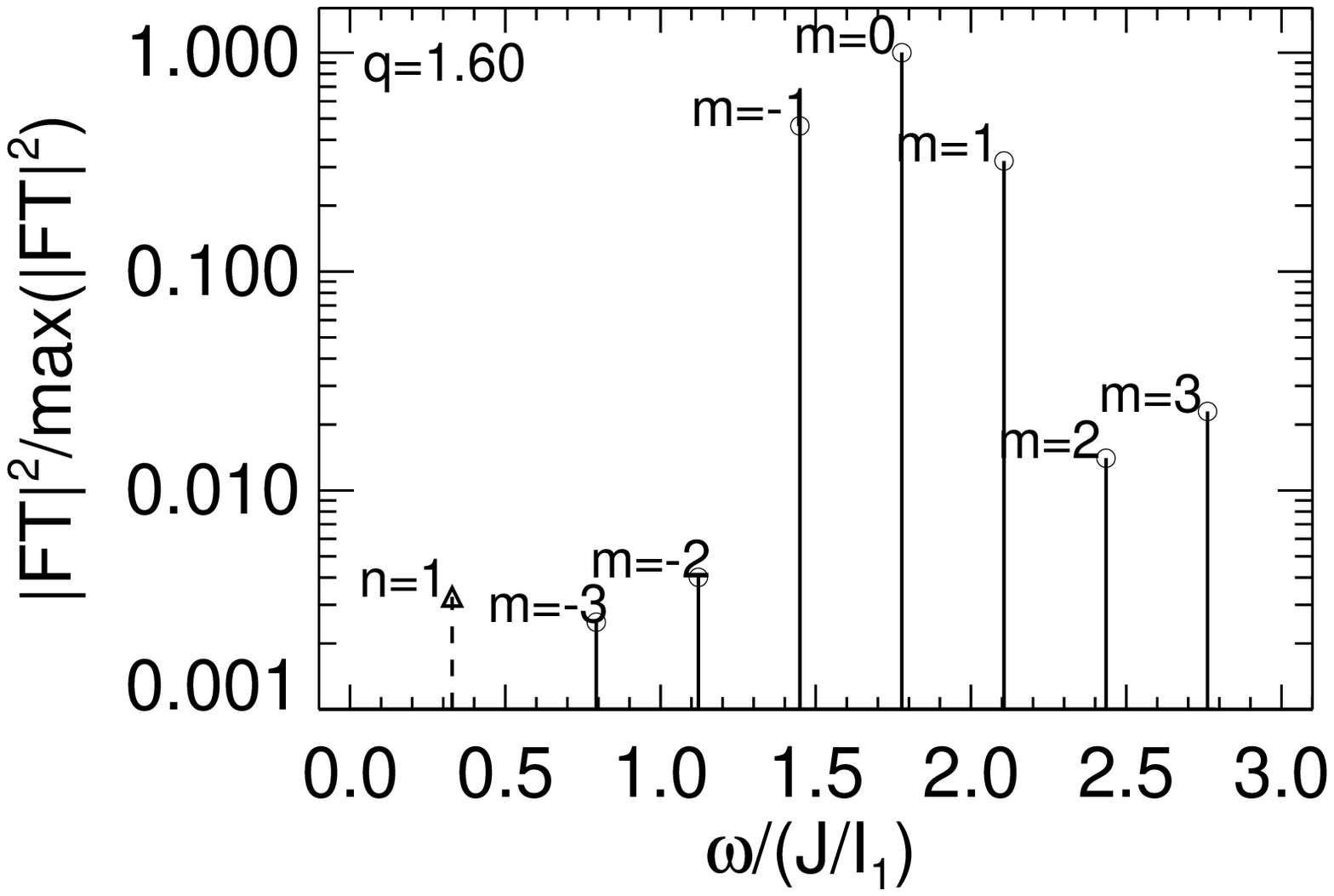}
\caption{Normalized power spectrum  of a torque-free rotating irregular grain
with $I_{1}:I_{2}:I_{3}=1:0.6:0.5$ 
for the different values of $q=1.05, 1.60$ 
(i.e. $q<q_{\sp}\equiv I_{1}/I_{2}$) and
$q=1.81>q_{\sp}$.
The components of $|{\FT}(\ddot{\mu}_x)|^2/\max(|{\FT}(\ddot{\mu}_x)|^2))$
(or $|{\FT}(\ddot{\mu}_y)|^2/\max(|{\FT}(\ddot{\mu}_x)|^2)$)
are indicated by circles, while the components of $|{\FT}(\ddot{\mu}_z)|^2/
\max(|{\FT}(\ddot{\mu}_x)|^2)$
are indicated by triangles. Orders of in-plane modes $m$  and out-of-plane modes 
$n$ are indicated, and case 1 ($\mu_{1}=\mu/\sqrt{3}$) of $\bmu$ orientation is assumed.
Figure reproduced from Hoang et al.~\cite{Hoang:2011p2326}.}
\label{fig:602}
\end{figure*}

Figure \ref{fig:602} presents normalized power spectra (squared amplitude of
Fourier transforms), $|{\FT}(\mu_{x,y})|^{2}/\max(|{\FT}(\mu_{x})|^{2})$ and 
$|{\FT}(\mu_{z})|^{2}/\max(|{\FT}(\mu_{x})|^{2})$
for the components $\ddot\mu_{x}$ (or $\ddot\mu_{y}$) and $\ddot{\mu}_{z}$, for 
a freely rotating irregular grain having the ratio of moments of inertia 
$I_{1}:I_{2}:I_{3}=1:0.6:0.5$ and for various $q$. 
Circles and triangles indicated with $m$ and $n$ denote peaks
of the power spectrum for oscillating components of $\ddot\mu_{x}$ (or $\ddot\mu_{y}$) and
$\ddot\mu_{z}$, respectively. The horizontal axis is the angular frequency
of emission modes normalized over the frequency of emission when
the grain spins around its shortest axis.

Multiple frequency modes are observed in the power spectra
of the irregular grain, but in Figure \ref{fig:602} we  show only 
the modes with power no less than $10^{-3}$ the maximum value.
One can see that for the case with large $q=1.6$, the modes with
$\omega/(J/I_{1})>1$ have power increasing, while the modes with
$\omega/(J/I_{1})<1$ have power decreasing. It indicates that
if grain rotational energy is increased so that the grain spends a
significant fraction of time rotating with large $q$, then the grain
should radiates larger rotational emission.

Although one should not expect the analytical expression
of power spectrum for the triaxial grain, the frequency modes
can be approximately found.
Indeed, for $q<I_{1}/I_{2}$, we found that power spectra for $\ddot\mu_{x}$ 
(or $\ddot\mu_{y}$) have angular frequency modes 
\bea
\omega_{m}\approx \langle \dot\phi\rangle+ 
m\langle|\dot\psi|\rangle,\label{eq:610}
\ena
where the bracket denotes the averaging value over
time, and $m=0,\pm 1,\pm 2...$ denote the order of the mode. 
The frequency modes for $\ddot\mu_{z}$  are given by 
\bea
\omega_{n}=n \langle|\dot{\psi}|\rangle,\label{eq:611}
\ena
where $n$ is integer and $n\ge 1$. 

In the following, the emission modes induced by the oscillation of $\mu_{x}$ or 
$\mu_{y}$, which lie in the $\xhat\yhat$ plane, perpendicular 
to $\bJ$, are called {\it in-plane} modes, and those induced by the oscillation of 
$\mu_{z}$ in the direction perpendicular to the $\xhat\yhat$ plane, are called
{\it out-of-plane} modes. The order of mode is denoted by
$m$ and $n$, respectively.
Figure \ref{fig:602} also shows that the emission power for out-of-plane modes 
$\omega_{n}$ is negligible compared to the power emitted by in-plane modes $\omega_{m}$.

Emission power spectra are numerically calculated to find $\omega_{k}$ and $P_{\ed, k}$, 
as functions of $J$ and $q$, for the various ratio of moments of inertia $I_{1}:I_{2}:I_{3}$. 
The obtained data will be used later to compute spinning dust emissivity.

\section{Grain Angular Momentum Distribution: Langevin Equation}
\subsection{Langevin Equation}
To find the exact distribution function for grain angular momentum
$\bJ$, Hoang et al.~\cite{Hoang:2010p2383} and Hoang et al.\cite{Hoang:2011p2326}
proposed a numerical approach based on the Langevin equation. Basically, they
numerically solved the Langevin equation describing the evolution of three
components of $\bJ$ in an inertial coordinate system. They read
\bea
dJ_{i}=A_{i}dt+\sqrt{B_{ii}}dq_{i},\mbox{~for~} i=\mbox{~x,~y,~z},\label{eq:612}
\ena
where $dq_{i}$ are random Gaussian variables with $\langle dq_{i}^{2}\rangle=dt$, 
$A_{i}=\langle {\Delta J_{i}}/{\Delta t}\rangle$ 
and $B_{ii}=\langle \left({\Delta J_{i}}\right)^{2}/{\Delta t}\rangle$ are 
damping and diffusion coefficients defined in the inertial coordinate system.
Detailed expressions of these coefficients can be found in Hoang et al.~\cite{Hoang:2010p2383}
and Hoang et al.~\cite{Hoang:2011p2326}.

For an irregular grain, to simplify calculations, we adopt the $A_{i}$ and
$B_{ii}$ for a disk-like grain obtained in HDL10. Following DL98b and HDL10, the disk-like
grain has the radius $R$ and thickness $L=3.35$ \Angstrom, and the ratio of moments of inertia along 
and perpendicular to the grain symmetry axis $h=I_{\|}/I_{\perp}$.
Thereby, the effect of nonaxisymmetry on $A_{i}$ and $B_{ii}$ is ignored, 
and we only examine the effect of grain wobbling resulting from the
grain triaxiality. 

In dimensionless units, $\bJ'\equiv \bJ/I_{\|}\omega_{\T,\|}$ with
$\omega_{\T,\|}\equiv \left(2k_{\B}T_{\gas}/I_{\|}\right)^{1/2}$ being the thermal
angular velocity of the grain along the grain symmetry axis,
and $t'\equiv t/\tau_{\H,\|}$, Equation (\ref{eq:612}) becomes 
\bea
dJ'_{i}=A'_{i}dt'+\sqrt{B'_{ii}}dq'_{i},\label{eq:613}
\ena
where $\langle dq_{i}^{'2}\rangle=dt'$ and
\bea
A'_{i}&=&-\frac{J'_{i}}{\tau'_{\gas,{\eff}}}-\frac{2}{3}\frac{J_{i}^{'3}}
{\tau'_{\ed,\eff}},~~~~\\
B'_{ii}&=&\frac{B_{ii}}{2I_{\|} k_{\B} T_{\gas}}\tau_{\H,\|},
\ena
where 
\bea
\tau'_{\gas,{\eff}}&=& \frac{\tau_{\gas,{\eff}}}{\tau_{\H,\|}}=
\frac{F_{\tot,\|}^{-1}}{\cos^{2}\theta+
\gamma_{\H}\sin^{2}\theta},\\
\gamma_{\H}&=&\frac{F_{\tot,\perp}\tau_{\H,\|}}{F_{\tot,\|}
\tau_{\H,\perp}},~
\tau'_{\ed,{\eff}}=\frac{\tau_{\ed,{\eff}}}{\tau_{\H,\|}},~~~
\ena
where $\tau_{\H,\|}$ and $\tau_{\H,\perp}$ are rotational damping times due to gas of 
purely hydrogen atom for rotation along parallel and perpendicular direction
to the grain symmetry axis $\ba_{1}$, 
$\tau_{\ed,\eff}$
is the effective damping time due to electric dipole emission (see HDL10, HLD11),
$\theta$ is the angle between $\ba_{1}$ and $\bJ$, and $F_{\tot,\|}$ and 
$F_{\tot,\perp}$ are total damping coefficients parallel and perpendicular 
to $\ba_{1}$ (see HDL10). In the case of fast internal relaxation, the diffusion
coefficients $A$ and $B$ are averaged over the distribution function $f_{\rm VRE}$

The Langevin equation (\ref{eq:613}) is solved using the numerical integration
with a constant timestep. 
At each timestep, the angular momentum $J_{i}$ obtained from LEs is recorded and
later used to find the distribution function $f_{J}$ with normalization
$\int_{0}^{\infty} f_{J}dJ=1$.

\subsection{Advantages of the Langevin Equation approach}
There are two apparent advantages of the LE approach. First, it allows us to treat
the spinning dust emission from grains with an arbitrary grain vibrational temperature.
Second, the impulsive excitation
by single-ion collisions, which can deposit an amount of angular momentum
greater than the grain angular momentum prior the collision, is
easily included in Equation (\ref{eq:612}) (see \cite{Hoang:2011p2326}).
Below, we briefly discuss the effect of impulsive excitations
arising from single-ion collisions.

DL98b showed that for grains smaller than $7\Angstrom$ the
angular impulse due to an individual ion-grain collision may be
comparable to the grain angular momentum prior the collision. Thus, infrequent hits
of ions can result in the transient rotational excitation for very small grains.

Let $\tau_{\icoll}^{-1}$ be the mean rate of ion collisions 
with the grain, given by
\bea \tau_{\icoll}^{-1}&=&f(Z_{\g}=0) n_{i}\pi
a^{2}\left(\frac{8k_{\B} T_{\gas}}{m_{i}\pi}\right)^{1/2}
\left[1+\frac{\sqrt{\pi}}{2}\Phi \right]+
\nonumber
\\ &&\sum_{Z_{\g}\ne
0}f(Z_{\g}) n_{\i}\pi
a^{2}\left(\frac{8k_{\B} T_{\gas}}{m_{i}\pi}\right)^{1/2}g
\left(\frac{Z_{\g}Z_{i}e^{2}}{ak_{\B} T_{\gas}}\right),~~~~~\label{eq:542}
\ena 
where $\Phi=\left({2Z_{i}^{2}e^{2}}/{ak_{\B} T_{\gas}}\right)^{1/2}$, $ g(x)=
1-x$ for $x<0$ and $g(x)=e ^{-x}$ for $x>0$, and
$f(Z_{\g})$ is the grain charge distribution function.
The probability of the next collision occurring in $[t,t+dt]$ is 
\bea dP=
\tau_{\icoll}^{-1}\exp\left(-t/\tau_{\icoll}\right)dt.\label{eq:dP}
\ena

The rms angular momentum per ion collision $\langle \delta J^{2}\rangle$ 
is inferred by dividing the total rms angular momentum to the collision rate, 
and its final formula is given in Hoang et al.~\cite{Hoang:2010p2383}.

Provided that the random moment of a single-ion collision is obtained from Equation (\ref{eq:dP}),
the angular momentum that the grain acquires through each single-ion collision can
easily be incorporated into the Langevin equation (\ref{eq:612}).
Hoang et al.~\cite{Hoang:2010p2383} found that the impulsive
excitations ions extend the distribution of grain angular momentum
to the region of high angular momentum (see next Section for
its effect on spinning dust emission).

\section{Spinning Dust Emissivity}
\subsection{Spinning grain of triaxial ellipsoid shape}
An irregular grain rotating with a given
angular momentum $J$ radiates at frequency 
modes $\omega_{k}\equiv \omega_{m}$ with $m=0,\pm 1, \pm 2...$ 
and $\omega_{k}\equiv \omega_{n}$ with $n=1,2,3...$ (see Equations \ref{eq:610} and 
\ref{eq:611}). For simplicity, let denote
the former
as $\omega_{m_{i}}$ and the latter as $\omega_{n_{i}}$ where 
$i$ indicates the value for $m$ and $n$. These frequency modes depend on 
the parameter $q(s)$, which is determined by the internal thermal fluctuations within the grain. 

To find the spinning dust emissivity by a grain at an observational frequency
$\nu$, first we need to know how much emission that is contributed by
each mode $\omega_{k}$.

Consider an irregular grain rotating with the angular momentum $J$, the probability 
of finding the emission at the angular frequency $\omega$ depends on the 
probability of finding the value $\omega$ such that
\bea
pdf(\omega|J)d\omega=f_{\VRE}(s,J)ds= A{\exp}\left(-\frac{q(s)J^{2}}
{2I_{1}k_{\B}T_{\vib}}\right)ds,~~~~~\label{eq:614}
\ena
where we assumed the VRE regime with $f_{\VRE}$ given by Equation (\ref{eq:606}).

For the mode $\omega\equiv\omega_{k}(s)$, from Equation (\ref{eq:614}) 
we can derive 
\bea
pdf_{k}(\omega|J)=\left(\frac{\partial \omega_{k}}{\partial s}\right)^{-1}
f_{\VRE}(s,J).
\ena

The emissivity from the mode $k$ is calculated as
\bea
j_{\nu,k}^{a}&=&\frac{1}{4\pi}\int_{J_{l}}^{J_{u}}P_{\ed,k}(J,q_{\le})f_{J}(J)
pdf_{k}(\omega|J)2\pi~ dJ\nonumber\\
&&+\frac{1}{4\pi}\int_{J_{l}}^{J_{u}}P_{\ed,k}(J,q_{>})f_{J}(J) 
pdf_{k}(\omega|J)2\pi~ dJ,~~~~~\label{eq:615}
\ena
where $q_{\le}$ and $q_{>}$ denote $q\le q_{\sp}$ and $q>q_{\sp}$, respectively;
$J_{l}$ and $J_{u}$ are lower and upper limits for $J$ 
corresponding to a given angular frequency $\omega_{k}(J,q)=\omega$, and 
$2\pi$ appears due to the change of variable from $\nu$ to $\omega$.

Emissivity by a grain of size $a$ at the observation frequency $\nu$ arising
from all emission modes is then
\bea
j_{\nu}^{a}&\equiv&\sum_{k}j_{\nu,k}^{a}.\label{eq:616}
\ena

Consider for example the emission mode $k\equiv \m_{0}$. For the case 
$I_{2}$ slightly larger than $I_{3}$, this mode has the angular frequency 
$\omega_{\m_{0}}=\langle \dot\phi\rangle=(J/I_{1})
q_{0}$ with $q_{0}$ obtained from  calculation of $\omega_{\m_{0}}$, is 
independent on $q$ for  $q<q_{\sp}$.\footnote{$q_{0}$ approaches $I_{1}/I_{2}$ 
as $I_{3} \rightarrow I_{2}$, i.e., when 
irregular shape becomes spheroid.}  As a result 
\bea
pdf_{\m_{0}}(\omega|J)=\delta\left(\omega-(J/I_{1})q_{0}\right).
\ena
Thus, the first term of Equation 
(\ref{eq:615}), denoted by $j_{\nu,\m_{0},\le}^{a}$, is rewritten as
\bea
j_{\nu,\m_{0},\le}^{a}&=&\frac{1}{2}\int_{J_{l}}^{J_{u}}P_{\ed,\m_{0}}
(J,q_{\le})f_{J}(J)
\delta \left(\omega-(J/I_{1}) q_{0}\right)dJ,\nonumber\\
&=&\frac{1}{2}\frac{I_{1}f_{J}(J_{0})}{q_{0}} P_{\ed,\m_{0}}(J_{0},q(s)),
\label{eq:617}
\ena
where $J_{0}=I_{1}\omega/q_{0}$, and the value of $q(s)$ remains to be determined.

For $q>q_{\sp}$, $\langle \dot\phi\rangle$ is a function of $q$. Hence, 
the emissivity (\ref{eq:615}) for the mode $k\equiv \m_{0}$ becomes
\bea
j_{\nu,m_{0}}^{a}&=&\frac{1}{2}\frac{I_{1}f_{J}(J_{0})}{q_{0}}
\int_{0}^{s_{\sp}} ds P_{\ed,\m_{0}}(J_{0},q(s))f_{\VRE}(J_{0},s)\nonumber\\
&&+\frac{1}{2}\int_{J_{l}}^{J_{u}}P_{\ed,\m_{0}}(J,q_{>})f_{J}(J) 
pdf_{\m_{0}}(\omega|J)dJ,~~~~~\label{eq:618}
\ena
where $s_{\sp}$ is the value of $s$ corresponding to $q=q_{\sp}$, and the term 
$P_{\ed,\m_{0}}(J_{0},q(s))$ in Equation (\ref{eq:617})
has been replaced by its average value over the internal thermal distribution $f_{\VRE}$.

The emissivity per H is obtained by integrating $j_{\nu}^{a}$ over the grain 
size distribution:
\bea
{j_\nu\over n_\H} = 
{1\over n_\H}
\int_{a_{\min}}^{a_{\max}} da {dn\over da}j_{\nu}^{a}~,\label{eq:619}
\ena
where $j_{\nu}^{a}$ is given by Equation (\ref{eq:616}).

\subsection{A degenerate case: grains of disk-like shape}
The spinning dust emissivity from disk-like grains (e.g., $I_{2}=I_{3}$)
is a degenerate case of triaxial grains. Basically,
a disk-like grain with an angular momentum $\bJ$ radiates at 
four frequency modes: 
\bea
\omega_{\m_{i}}&\equiv&\dot\phi+i \dot\psi=\frac{J}{I_{\|}}
\left[h+i(1-h)\cos\theta\right],\label{eq:619a}\\
\omega_{n_{1}}&\equiv&\dot\psi=\frac{J}{I_{\|}}(1-h)\cos\theta,\label{eq:620}
\ena
where $i=0$ and $\pm 1$ (see HDL10; \cite{AliHaimoud:2009p641}).

The emission power of these modes are given by following analytical forms 
(HDL10; \cite{Silsbee:2011p5567}):
\bea
P_{\omega_{\m_{0}}}&=&\frac{2\mu_{\|}^{2}}{3c^{3}}\omega_{\m_{0}}^{4}
\sin^{2}\theta,
\label{eq:621}\\
P_{\omega_{\m_{\pm1}}}&=&\frac{\mu_{\perp}^{2}}{6c^{3}}\omega_{\m_{\pm1}}^{4}
(1\pm\cos\theta)^{2},\label{eq:622}\\
P_{\omega_{n_{1}}}&=&\frac{2\mu_{\perp}^{2}}{3c^{3}}\omega_{n_{1}}^{4}
\sin^{2}\theta.
\label{eq:623}
\ena

For the disk-like grain, from Equation (\ref{eq:604}), the number of states
in phase space $s$ for $q$ spanning from $1-q$ becomes
\bea
s=1-\left(\frac{h-q}{h-1}\right)^{1/2}=1-\cos\theta,\label{eq:624}
\ena
where $q=1+(h-1)\sin^{2}\theta$ has been used.
Thus, for an arbitrary mode with frequency $\omega_{k}$, we obtain
\bea
pdf_{k}(\omega|J)d\omega=f_{\VRE}(s,J)ds=f_{\VRE}(\theta,J)\sin\theta d\theta.~~~
\ena

Taking use of $\omega=\omega_{k}(J,\theta)$, we derive
\bea
pdf_{k}(\omega|J)=f_{\VRE}(\theta,J)\left(\frac{\partial \omega_{k}}
{\partial \theta}\right)^{-1}\sin\theta.\label{eq:625}
\ena

Therefore, by substituting Equations (\ref{eq:621})-(\ref{eq:623}) in Equation 
(\ref{eq:615}), the emissivity at the observation frequency $\nu=\omega/(2\pi)$ from a 
disk-like grain of size $a$ is now given by
\bea
j_{\nu}^{a}&\equiv&\frac{1}{2} \frac{f_{J}(I_{\|}\omega/h)}{h}
\frac{2\mu_{\|}^{2}}{3c^{3}}\omega^{4}\langle\sin^{2}\theta\rangle\nonumber\\
&&+\frac{1}{2} \frac{\mu_{\perp}^{2}}{6c^{3}}\omega^{4}\int_{J_{l}}^{J_{u}}
pdf_{\m_{1}}(\omega|J)
f_{J}(J)dJ\nonumber\\
&&+\frac{1}{2} \frac{\mu_{\perp}^{2}}{6c^{3}}\omega^{4}\int_{J_{l}}^{J_{u}}
pdf_{\m_{-1}}(\omega|J)
f_{J}(J)dJ\nonumber\\
&&+\frac{1}{2} \frac{\mu_{\perp}^{2}}{3c^{3}}\omega^{4}\int_{J_{l}}^{J_{u}}
pdf_{n_{1}}(\omega|J)
f_{J}(J)dJ,\label{eq:626}
\ena
where $pdf_{m_{\pm1}}$ and $pdf_{n_{1}}$ are easily derived by using 
Equation (\ref{eq:625}) for $\omega_{m_{\pm1}}$ and $\omega_{n_{1}}$,
and $J_{l}=I_{\|}\omega/(2h-1)$ and $J_{u}=I_{\|}\omega$ for $m_{\pm 1}$ mode, 
and $J_{l}=I_{\|}\omega/(h-1)$ and $J_{u}=\infty$ for $n_{1}$ mode (see Equations 
\ref{eq:619a} and \ref{eq:620}) .

\subsection{Emissivity}

Hoang et al.~\cite{Hoang:2011p2326}
assumed that smallest grains of size $a \le a_{2}=6 \Angstrom$,
have irregular shape, and larger grains are spherical. To compare the emissivity
from an irregular grain with that from
a disk-like grain, they considered the simplest case of the irregular shape in which
the circular cross-section of the disk-like grain is adjusted to the elliptical 
cross-section. The emission by two grains of different shapes with the same
mass $M$ and thickness $L$ is under interest, therefore, the semi-axes of the elliptical disk is constrained by
the grain mass:
\bea
M=\pi R^{2}L=\pi b_{2}b_{3}L,\label{eq:629}
\ena
where $R=(4a^{3}/3L)^{1/2}$ is the radius of the disk-like grain,
$b_{2}$ and $b_{3}$ 
are the length of semi-axes $\ba_{2}$ and $\ba_{3}$, and $b_{1}=L$
is kept constant. Assuming that the circular disk is 
compressed by a factor $\alpha\le 1$ along $\ba_{2}$, then Equation (\ref{eq:629})
yields
\bea
b_{2}=\alpha R,~~b_{3}=\alpha^{-1}R.
\ena
Denote the parameter $\eta\equiv b_{3}/b_{2}=\alpha^{-2}$, then the degree of grain 
shape irregularity is completely characterized by $\eta$.

For each grain size $a$, the parameter $\eta$ is increased from $\eta=1$ 
to $\eta=\eta_{\max}$. However, $\eta_{\max}$ is constrained by the fact that
the shortest axis $\ba_{2}$ should not be shorter than the grain thickness $L$.
The value $\eta_{\max}\sim 3/2$ is conservatively chosen.

Although the irregular grain can radiate at a large number of frequency modes,
only the modes with the order $|m| \le 2$ are important. The higher order modes
contribute less than $\sim 0.5\%$ to the total emission, thus they are neglected.
Hoang et al.~\cite{Hoang:2011p2326} assumed that grains smaller than $a_{2}$ have
a fixed vibrational temperature $T_{\vib}$ (see Hoang et al.~\cite{Hoang:2011p2326} for the detailed
treatment of $T_{\vib}$ distribution), and that for the instantaneous
value of $J$ the rotational energy has a probability distribution $f_{\VRE}$
(i.e. VRE regime, see Eq.\ref{eq:606}).

The grain size distribution $dn/da$ from Draine \& Li~\cite{Draine:2007p4735}
is adopted with the total to selective extinction $R_{V}=3.1$ and the total carbon
abundance per hydrogen nucleus $b_{\rm C}=5.5\times 10^{-5}$ in carbonaceous 
grains with $a_{\min}=3.55 \Angstrom$ and $a_{\max}=100 \Angstrom$.

The spinning dust emissivity is calculated for a so-called model A (similar
to DL98b; HDL10), in which $25\%$ of
grains have the electric dipole moment
parameter $\beta=2\beta_{0}$, $50\%$ have $\beta=\beta_{0}$ and $25\%$ have 
$\beta=0.5\beta_{0}$ with $\beta_{0}=0.4$ D. In the rest of the paper, the 
notation {\it model} A is omitted, unless stated otherwise.

\begin{figure*}
\includegraphics[width=0.5\textwidth]{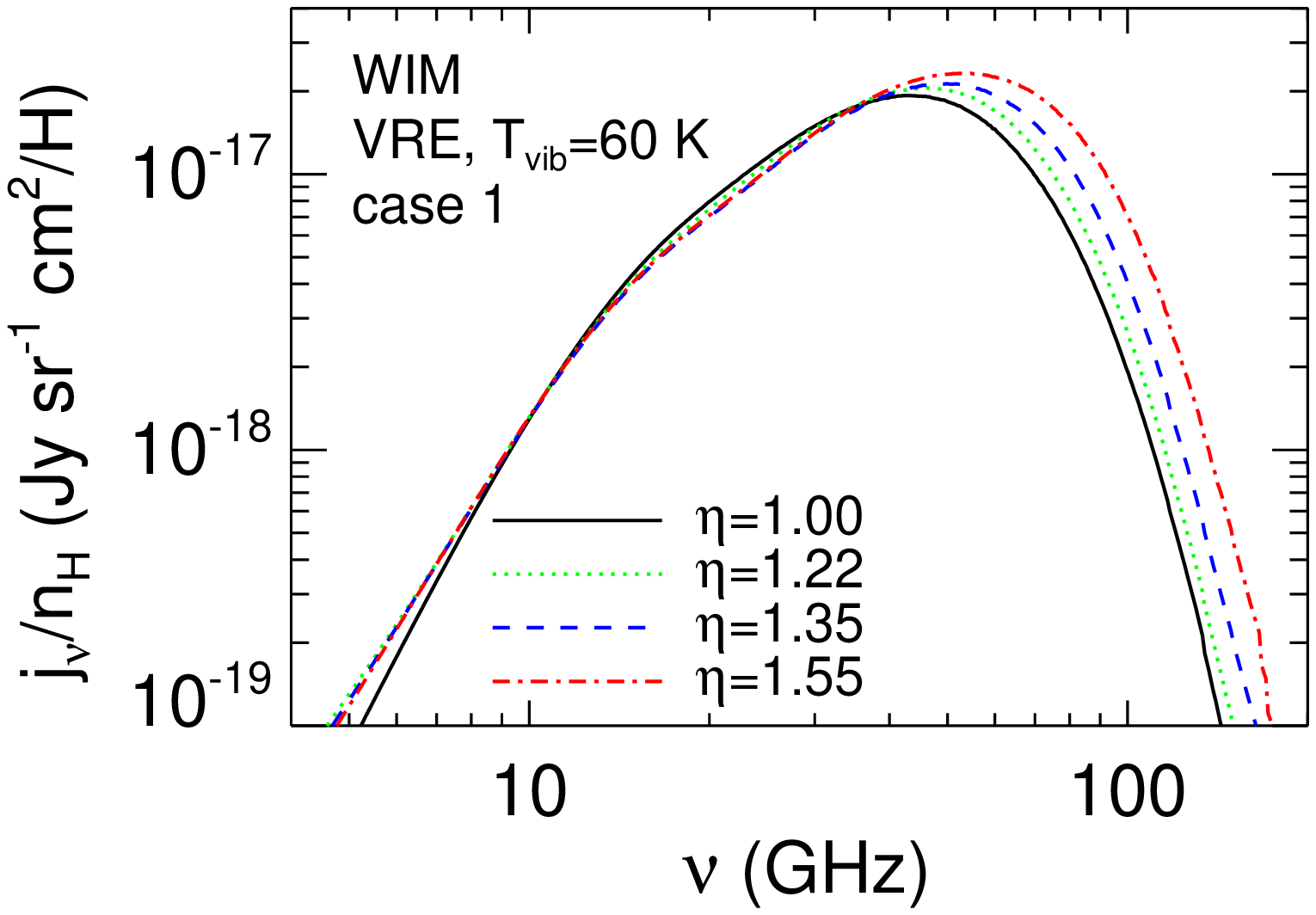}
\includegraphics[width=0.5\textwidth]{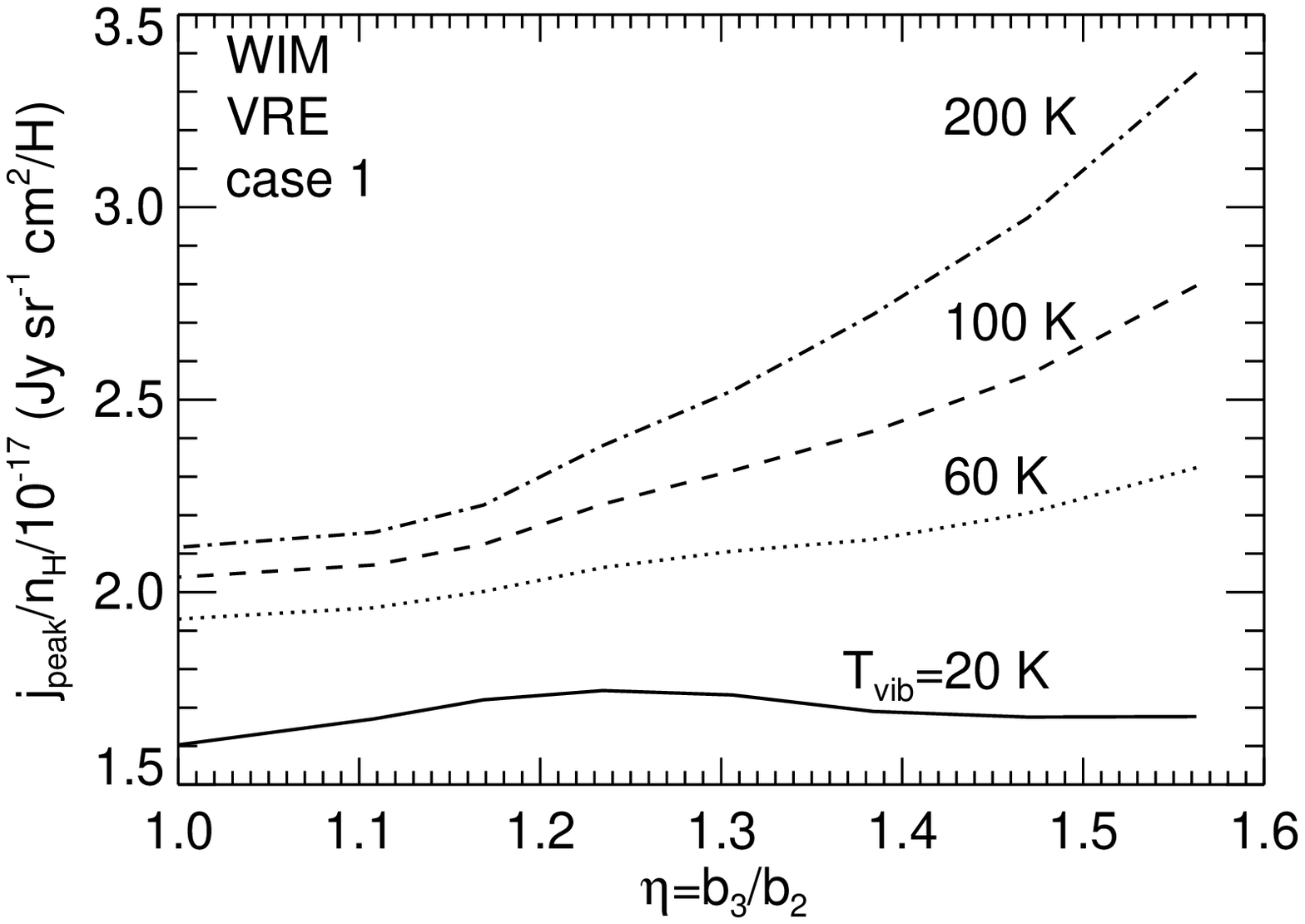}
\caption{Emissivity per H from irregular grains of different degrees of
irregularity $\eta=b_{3}/b_{2}$  with $T_{\vib}=60$
and $200$ K in the WIM for the case in which the electric dipole moment
is isotropically oriented in the grain body (i.e., case 1 with $\mu_{1}=\mu/\sqrt{3}$).
The emission spectrum shifts to higher frequency as $\eta$ decreases (i.e., grain becomes
more irregular). Here the grain mass is held fixed as $\eta$ changes.
Figure reproduced from Hoang et al.~\cite{Hoang:2011p2326}.}
\label{fig:604}
\end{figure*}

The left panel in Figure \ref{fig:604} shows the spinning dust emissivity for different degrees of
irregularity $\eta$ and with a dust temperature $T_{\vib}=60\K$ in the WIM.
The emission spectrum for a given $T_{\vib}$ shifts to higher frequency as
$\eta$ decreases (i.e. the degree of  grain irregularity increases),
but their spectral profiles remain similar. The right panel shows the
increase of peak emissivity $J_{\peak}$ with increasing $\eta$.

One particular feature in Figure \ref{fig:604}{\it right} is that
for axisymmetric grains ($\eta=1$), the emissivity increases by
a factor of $1.3$ with $T_{\vib}$ increases from $20-200\K$.
However, for the irregular grain with high triaxiality $\eta=1.5$,
the emissivity increases by a factor of $2$. The peak frequency
is increased by a factor of $1.4$.

This feature is easy to understand because the irregular grain
radiate at more frequency modes than the axisymmetric grain.
As a result, for the grain temperature increases to a sufficiently
high value, it results in the uniform distribution of the angle
between grain symmetry axis and angular momentum,
so that the spinning dust emissivity becomes saturated.
On the other hand, for the triaxial grain, as $T_{\vib}$
increases, it allows grain to rotate about its axis of minimum
inertia (smallest moment of inertia). As a result, the grain
radiates at frequency modes with higher frequency,
and power.

In the case of efficient IVRET, vibrational energy is converted to rotational
emission, which results in the increase of both emissivity and peak frequency.
As shown, the energy transfer is more efficient for more irregular
grain. The reason for that is, the more irregular grain allows
grain spends a larger fraction of time rotating along the axis of minor inertia. 

\begin{figure}
\begin{center}
\includegraphics[width=0.8\textwidth]{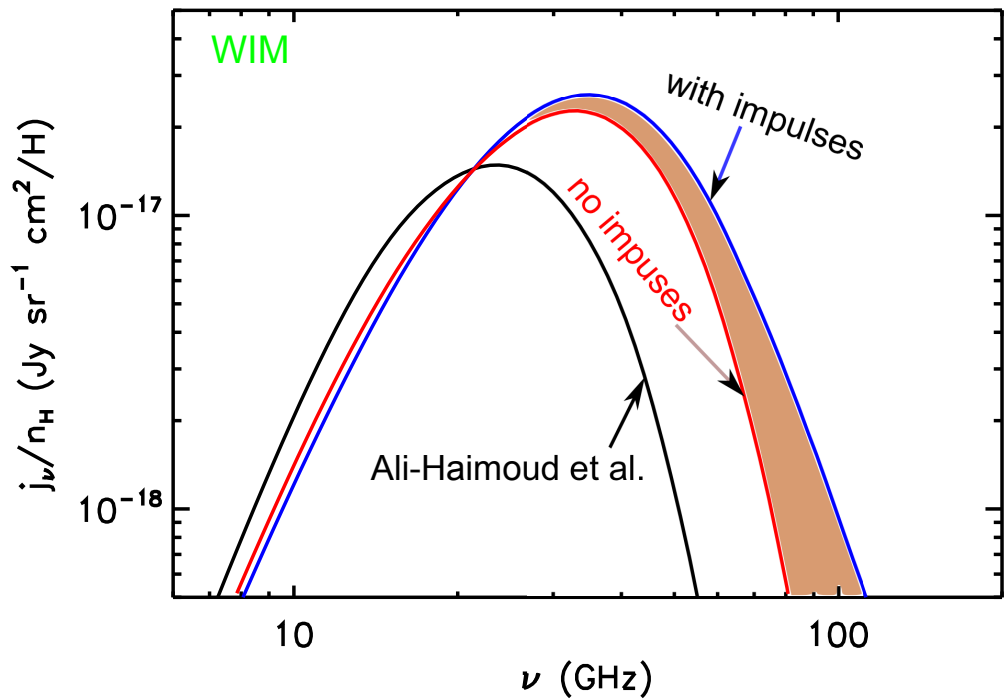}
\caption{Emissivity per H obtained for WIM
without ionic impulses using the Fokker-Planck equation from Ali-Ha{\"\i}moud et al.~\cite{AliHaimoud:2009p641} and with impulses
using our LE simulations for grain wobbling. The spectra are
efficiently broadened as a result of impulses (see blue line). Figure reproduced from Hoang et al.~\cite{Hoang:2010p2383}.}
\label{f516}
\end{center}
\end{figure}

The effect of impulsive excitations by single-ion collisions
is shown in Figure \ref{f516}. One can see that the impulses from ions 
can increase the emissivity by $\sim 23\%$, and slightly increase
the peak frequency (see Figure \ref{f516}). The tail of high frequency
part is obviously extended due to the contribution from ionic impulses
with large angular momentum.

\section{Constraining spinning dust parameters and implications}

Spinning dust emission involves a number of parameters, including
grain physical parameters and environment parameters.
Among them, the grain dipole moment and gas density are two
most important parameters, but they can be constrained
using theoretical modeling combined with observation data
(see e.g., Dobler et al.~\cite{2009ApJ...699.1374D}; Hoang et al.~\cite{Hoang:2011p2326}).
In the following, we discuss a number of parameters, which are
shown to be important but more difficult to constrain through observation.

\subsection{Lower cutoff of grain size distribution $a_{\min}$}\label{sec:amin}

The spinning dust emission spectrum is sensitive to the population of tiny dust grains,
and its peak frequency is mostly determined by
the smallest PAHs. Let $a_{\min}$ be the size of smallest PAHs. When $a_{\min}$ is increased, the peak
frequency $\nu_{\peak}$ decreases accordingly.

\begin{figure}
\begin{center}
\includegraphics[width=0.8\textwidth]{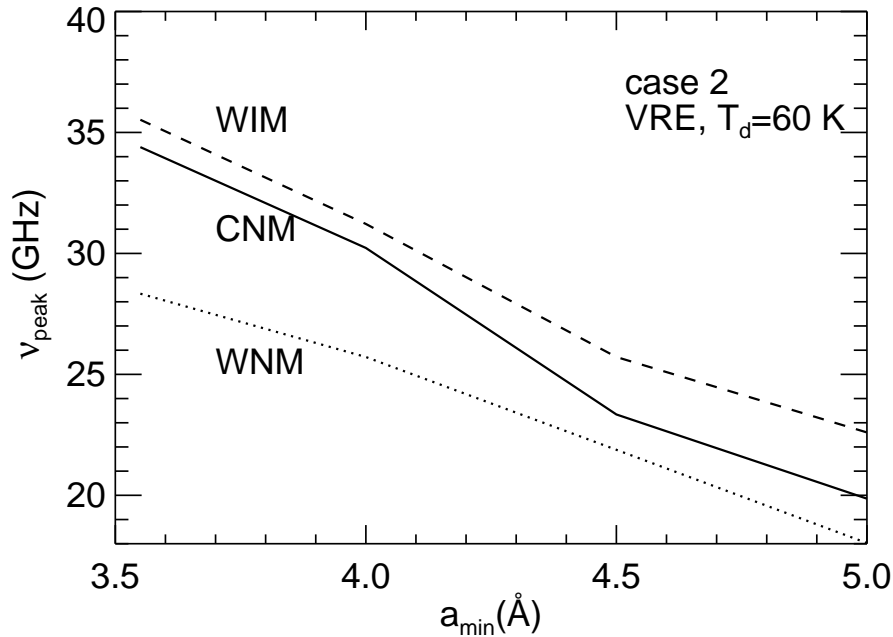}
\caption{Decrease of the peak frequency $\nu_{\peak}$ of spinning dust spectrum 
with the lower cutoff of grain size distribution $a_{\min}$ for
various environment conditions. Figure reproduced from Hoang et al.~\cite{Hoang:2011p2326}.}
\label{fig:608}
\end{center}
\end{figure}

Figure \ref{fig:608} shows the variation of $\nu_{\peak}$ as 
a function of $a_{\min}$ for various environments,
for the case in which the grain dipole moment lies in
the grain plane (case 2) with $\mu_{1}=0$ and with the VRE regime ($T_{\rm d}=60~\K$). 
As expected, $\nu_{\peak}$ decreases generically with  $a_{\min}$ increasing.
Thus, in addition to grain dipole moment, the lower cutoff of grain
size also plays an important role.

\subsection{Constraining the shape of very small grains}

Very small grains and PAHs are expected to be nonspherical. However, contraining
grain triaxiality using spinning dust appears rather challenging. In the simplest
case where the grain shape can be approximated as an triaxial ellipsoid, the
possibility is still low because there are many parameters involved in spinning dust.

\subsection{Can compressible turbulence be observed through spinning dust emission?}

The discussion of interstellar conditions adopted in DL98 and other works
on spinning dust was limited by idealized interstellar phases.
It is now recognized that turbulence plays
an important role in shaping the interstellar medium.

For spinning dust, the turbulence can increase the emissivity
due to its nonlinear dependence on material density.
Indeed, in a medium with density fluctuations, the effective emissivity is 
\bea
\langle j_{\nu}\rangle=\int_{0}^{1} f(x) j_{\nu}(x\langle\rho\rangle)dx,
\ena
where $f(x)dx$ is the fraction of the mass with $\rho/\langle\rho\rangle \in
(x,x+dx)$. We use compression distributions $f(x)$ obtained from MHD simulations
for $M_{\s}=2$ and $7$ to evaluate $\langle j_{\nu}\rangle$ for 
the WIM and CNM, respectively. 

\begin{figure}
\begin{center}
\includegraphics[width=0.8\textwidth]{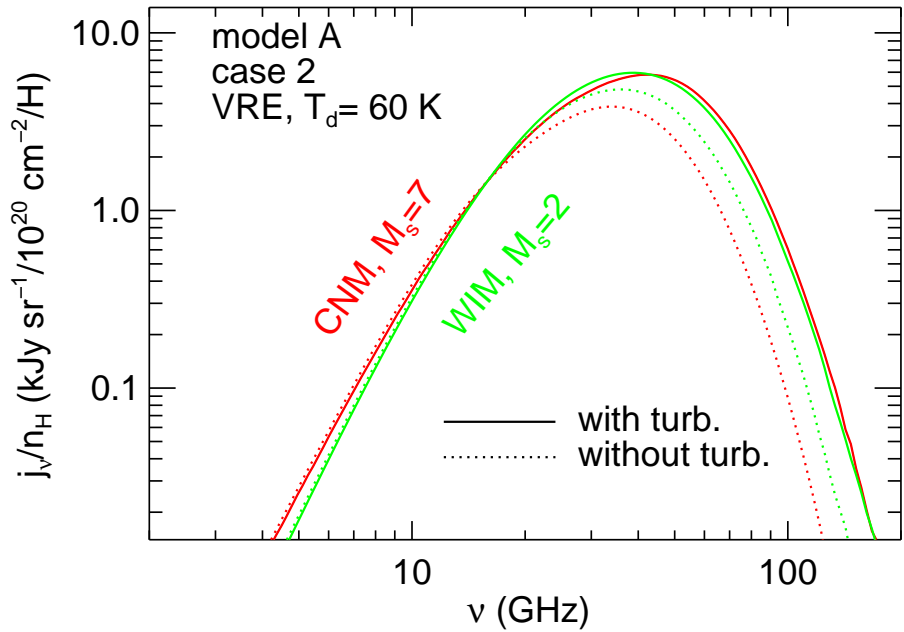}
\caption{Spinning dust emissivity per H 
in the presence of compressible turbulence 
with sonic Mach number $M_{\s}=2$ and $7$, compared to that from 
uniform medium with $n_{\H}=\overline{n}_{\H}$ for the CNM (red) and WIM (green). 
The peak emissivity is increased, and the spectrum is shifted to 
higher frequency due to compressible turbulence. Case 2 ($\mu_{1}=0$) of $\bmu$ orientation
 is considered. Figure reproduced from Hoang et al.~\cite{Hoang:2011p2326}.}
\label{fig:612}
\end{center}
\end{figure}

We assume the case 2 ($\mu_{1}=0$) of $\bmu$ orientation. The resulting effective emissivity
is compared with the emissivity from the uniform medium in Figure \ref{fig:612}.
It can be seen that the turbulent compression increases the emissivity,
and shifts the peak to higher $\nu_{\peak}$. The increase of emissivity is
significant for strong turbulent medium.

The distribution of phases, for instance, CNM and WNM of the ISM
at high latitudes can be obtained from absorption lines. Similarly, studying 
fluctuations of emission it is possible to constrain parameters of turbulence. 
In an idealized case of a single phase medium with fluctuations of density with
a given characteristic size one can estimate the value of the 3D fluctuation by
studying the 2D fluctuations of column density. A more sophisticated techniques
of obtaining sonic Mach numbers\footnote{It may be seen that Alfven Mach 
numbers have subdominant effect on the distribution of densities 
(see Kowal et al.~\cite{Kowal:2007p7242}). Thus in our study we did not vary the 
Alfven Mach number.} have been developed recently (see Kowal et al.~\cite{Kowal:2007p7242};
Esquivel \& Lazarian~\cite{Esquivel:2010p7246}; Burkhart et al.~\cite{Burkhart:2009p7249}). In particular,
Burkhart et al.~\cite{Burkhart:2010p3657}, using just column density 
fluctuations of the SMC, obtained a distribution of Mach numbers corresponding 
to the independent measurements obtained using Doppler shifts and absorption data. With
 such an input, it is feasible to quantify the effect of turbulence in actual 
observational studies of spinning dust emission.

\subsection{Effect of dust acceleration on spinning dust emission}
Collisions of ultrasmall grains with ions and neutrals in plasma appear to be a
dominant mechanism of rotational excitations
for spinning dust emission, particularly, in dark clouds where UV photons
are blocked out.
Current spinning dust models assume Brownian motion of grains
relative to gas, but it is known that
grains may move with suprathermal velocities due to acceleration
by turbulence (see e.g., \cite{1985prpl.conf..621D}; Yan \& Lazarian~\cite{2003ApJ...592L..33Y};
\cite{2004ApJ...616..895Y}; Hoang et al.~\cite{2012ApJ...747...54H}), and random charge
fluctuations (Ivlev et al.~\cite{2010ApJ...723..612I}; Hoang \& Lazarian~\cite{2011arXiv1112.3409H}).
The latter mechanism, namely random charge fluctuations-induced-acceleration,
is found to be efficient for tiny grains (Hoang \& Lazarian~\cite{2011arXiv1112.3409H}).

The resonant acceleration by fast modes of MHD turbulence, which occurs when
the grain gyroradius is comparable to the scale of turbulence eddy (i.e., $r_{g}\sim k^{-1}$)
is considered a dominant mechanism
for large grains ($>10^{-5}\cm$), whereas it is negligible for ultrasmall grains
because the grain gyroradius falls below the cutoff scale of the turbulence
due to viscous damping (see Yan et al.~\cite{2004ApJ...616..895Y};
Hoang et al.~\cite{2012ApJ...747...54H}).

In highly ionized media (e.g., WIM, HII regions) the resonant acceleration
by MHD turbulence may become important for ultrasmall grains because the damping cutoff of
MHD turbulence is suppressed due to the decrease of viscous neutral damping.
We also note that recent observations by Paladini et al.~\cite{Paladini:2012um})
revealed that PAHs and ultrasmall grains
may be present in HII regions, as shown through their $8\mu$m and $24\mu$m emission features,
respectively. 

Thus, assuming that grain rotational
kinetic energy is equal to its translational energy, the acceleration by these aforementioned
processes is expected to increase spinning dust emission. Further studies
should take into this issue into account.

\section{Polarization of Spinning Dust Emission and alignment of ultrasmall grains}
\subsection{Polarization of anomalous microwave emission}
Spinning dust emission is an important foreground component that
contaminates with the CMB radiation in the frequency $10-90$ GHz. The understanding
of how much polarized this emission component is becoming an pressing question
for future CMB B-mode missions. 

Recent observational
studies (Dickinson et al. \cite{2011MNRAS.418L..35D}; L\'{o}pez-Caraballo et al.~\cite{LopezCaraballo:2011p508}; Macellari et al. \cite{2011MNRAS.418..888M})
showed that the average polarization of AME is
between $2-5\%$. In the last several years, significant progress
has been made in understanding spinning dust emission, both theory and
observation; but the principal mechanism of alignment of ultrasmall grains
is not well understood.

\subsection{Alignment of ultrasmall dust grains}

Grain alignment is an exciting problem (see Lazarian~\cite{2007JQSRT.106..225L} for a review). 
The most promising mechanism for the grain alignment is based on radiative torques. Proposed
originally by Dolginov \& Mitrofanov~\cite{1976Ap&SS..43..291D} it is related to the interaction of
unpolarized radiation with {\it irregular} grains. The numerical studies in
Draine \& Weingartner~\cite{1996ApJ...470..551D} and \cite{1997ApJ...480..633D}
showed the efficiency and promise of the radiative torques (which later were termed RATs). The physical picture of the RAT alignment and a detailed study of important relevant effects is presented in Lazarian \& Hoang~\cite{2007MNRAS.378..910L}, \cite{2008ApJ...676L..25L} and Hoang \& Lazarian~
\cite{2008MNRAS.388..117H}, \cite{2009ApJ...697.1316H}, \cite{2009ApJ...695.1457H}. However, the efficiency of RATs plummets as the size of grains gets much smaller than the
radiation wavelength. Therefore, this mechanism, which seem to provide a good
correspondence with the optical and infrared data (see
Lazarian~\cite{2007JQSRT.106..225L}, Whittet et al.~\cite{2008ApJ...674..304W})
cannot be applicable to ultrasmall spinning dust.   

Microwave emission from spinning grains is expected to be polarized if
grains are aligned. Alignment of ultrasmall grains 
(essentially PAHs) is likely to be different from alignment
of large (i.e., $a>10^{-6}$~cm) grains discussed above.
One of the mechanisms that might produce the alignment of the ultrasmall
grains is the
paramagnetic dissipation mechanism proposed by Davis \& Greenstein~\cite{Davis:1951p3421}.
The Davis-Greenstein alignment mechanism (Davis \& Greenstein 1951, 
Roberge \& Lazarian 1999) is
straightforward: for a spinning grain
the component of interstellar magnetic field
perpendicular to the grain angular velocity varies in grain coordinates,
resulting in time-dependent magnetization, associated
energy dissipation, and a torque acting on the grain.
As a result grains tend to rotate with angular momenta parallel to the
interstellar magnetic field.

Lazarian \& Draine~\cite{Lazarian:2000p3023} (henceforth LD00) found
that the traditional picture of paramagnetic relaxation is
incomplete, since it
disregards the so-called ``Barnett magnetization'' (Landau \& Lifshitz~\cite{1960ecm..book.....L}).
The Barnett effect--the inverse of the Einstein-De Haas effect--
consists of the spontaneous magnetization of
a paramagnetic body rotating in field-free space. This effect can be understood in
terms of the lattice sharing part of its angular momentum with
the spin system. Therefore the implicit assumption in Davis \& Greenstein~\cite{Davis:1951p3421}
that the magnetization within a {\it rotating grain} in a {\it static}
magnetic field is equivalent to the magnetization within a
{\it stationary grain} in a {\it rotating} magnetic field is clearly not exact.

\begin{figure}
\begin{center}
\includegraphics[width=0.8\textwidth]{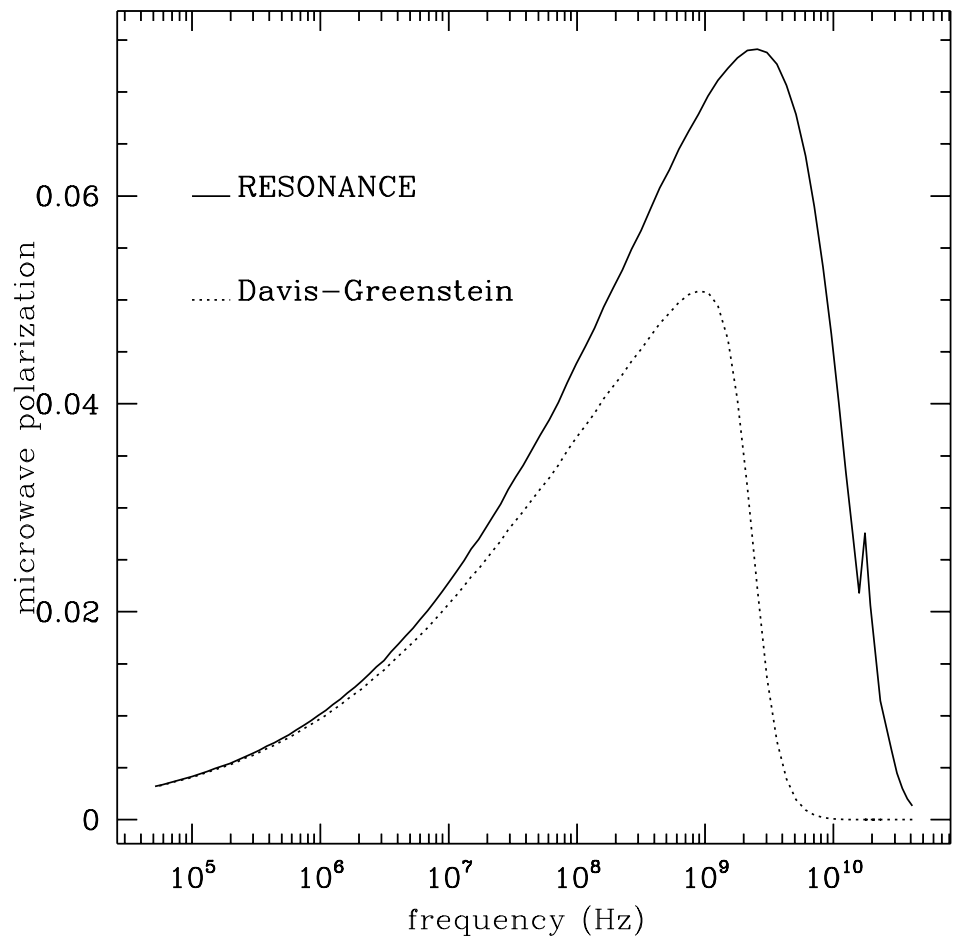}	
\caption{Polarization for both
resonance paramagnetic relaxation and Davis-Greenstein relaxation for grains in
the cold interstellar medium as a function of frequency (from LD00).
For resonance relaxation the saturation effects are neglected,
which means that the upper curves correspond to the {\it maximal}
values allowed by the resonant paramagnetic mechanism.}
\label{fig3}
\end{center}
\end{figure}

LD00 accounted for the ``Barnett magnetization'' and termed the effect
of enhanced paramagnetic relaxation arising from grain magnetization ``resonance
paramagnetic relaxation''. It is clear from Figure \ref{fig3} that resonance
paramagnetic relaxation persists
at the frequencies when the Davis-Greenstein relaxation vanishes. However
the polarization is marginal for $\nu>35$~GHz anyhow. The discontinuity
at $\sim 20$~GHz
is due to the assumption that smaller grains are planar, and larger
grains are spherical. The microwave emission will be polarized
in the plane perpendicular to magnetic field because the angular momentum
is partially aligned with the magnetic field.

\subsection{Constraining the alignment of ultrasmall grains}

{\it Can we constrain the alignment of ultrasmall grains through polarization of
mid-infrared ($2-12 \mu$m) emission features?}\\
The answer to this question is ``probably not''. Indeed, as discussed earlier,
mid-infrared emission from ultrasmall grains,
takes place as they absorb UV photons. These photons raise
grain vibrational temperature, randomizing grain axes in relation to
its angular momentum (see Lazarian \& Roberge~\cite{Lazarian:1997p2597}). Taking values
for Barnett relaxation from Lazarian \& Draine~\cite{Lazarian:1999p783}, we estimate
the randomization time of the $10^{-7}$~cm grain to be
$2\times 10^{-6}$~s, which is less than the grain cooling time due to IR emission. As a
result, the emanating infrared emission will be polarized very marginally.
If, however, Barnett relaxation is suppressed, the randomization time
is determined
by inelastic relaxation (Lazarian \& Efroimsky \cite{1999MNRAS.303..673L}) and is
$\sim 0.1$~s, which would entail a partial polarization of
infrared emission.

{\it Can we constrain the alignment of ultrasmall grains via the ultraviolet polarization?}\\
PAHs and ultrasmall grains that produce spinning dust emission
are likely the same particles that produce the prominent UV absorption feature at $2175$\AA~
(see e.g., Draine \& Li~\cite{2007ApJ...657..810D}). The lack of polarization excess at $2175$\AA~
is consistent with the expectation that the PAHs are poorly aligned.
However, the small degree of polarization (see Wolf et al.~\cite{1997ApJ...478..395W})
indicates that there must be some residual alignment of ultrasmall grains.
The constraint for such a residual alignment can be obtained by fitting theoretical
model with the UV polarization of starlight (Martin~\cite{2007EAS....23..165M}).
When the residual alignment available, one can predict the polarization level
of spinning dust. 

Apart from emission from spinning dust, another new type of emission
from dust is possible. Draine \& Lazarian~\cite{1999ApJ...512..740D} noticed that
the strongly magnetized material is capable of producing much
more microwave thermal emission compared with non-magnetic grains.
They suggested this as a possible alternative to spinning dust emission,
which can be responsible for a part or even most of the anomalous 
microwave emission. Such an emission can be strongly polarized, making
anomalous emission an important contaminant in terms of CMB polarization studies.

Further research showed that at the frequencies 20--90 GHz the spinning
dust dominates. However, Draine \& Hensley~\cite{Draine:2012zu} performed new calculations 
of microwave response of strongly magnetic grains. At higher frequencies this
new extensive study of evaluating microwave emissivity of strongly 
magnetic grains showed that magneto-dipole response of interstellar dust
may be extremely important.

\section{Summary}

The principal points discussed above are as follows:

\begin{itemize}

\item The model of spinning dust emission proposed by DL98 proved capable of explaining anomalous
microwave emission, and its predictions were confirmed by numerous observations
since the introduction of the model.

\item The DL98 spinning dust model has been improved recently by including the effects of
thermal fluctuations within dust grains, impulsive excitations with single ions, transient heating by UV
photons, triaxiality of grain shape, and compressible turbulence, which made the
spinning dust model more realistic.

\item Spinning dust emission involves a number of grain physical parameters and
environment parameters. With the latest progress on theoretical
modeling and observations, the possibility of using spinning dust as a diagnostics tool
for physical parameters of ultrasmall dust is open.

\item The spinning dust emission is expected to be partially polarized, but further
studies on alignment of ultrasmall grains and modeling of spinning dust polarization
is vitally required.

\end{itemize}

AL acknowledges the support of the Center for Magnetic Self-Organization
and the NASA grant NNX11AD32G. TH is grateful to Peter Martin for fruitful discussions
on constraining the alignment of ultrasmall grains and polarization
of anomalous microwave emission.

\end{document}